# Repurposing an Energy System Optimization Model for Seasonal Power Generation Planning

A.R. de Queiroz[a,b], D. Mulcahy[c], A. Sankarasubramanian[b], J.P. Deane[d], G. Mahinthakumar[b], N. Lu[c], J.F. DeCarolis[b]

[a] Department of Decision Sciences, School of Business
NC Central University
1801 Fayetteville Rd
Durham, NC 27707
Email: adequeiroz@nccu.edu

[b] Department of Civil, Construction, and Environmental Engineering
NC State University
2501 Stinson Drive Box 7908
Raleigh, NC 27695
Emails: arqueiroz@ncsu.edu, jdecarolis@ncsu.edu, sankar_arumugam@ncsu.edu, gmkumar@ncsu.edu

[c] Department of Electrical and Computer Engineering
NC State University
890 Oval Drive
Raleigh, NC 27606
Emails: djmulcah@ncsu.edu, nlu2@ncsu.edu

[d] Energy Policy and Modelling Group
University College Cork
College Road, Cork T12 K8AF
Email: Jp.Deane@ucc.ie

**Abstract**
Seasonal climate variations affect electricity demand, which in turn affects month-to-month electricity planning and operations. Electricity system planning at the monthly timescale can be improved by adapting climate forecasts to estimate electricity demand and utilizing energy models to estimate monthly electricity generation and associated operational costs. The objective of this paper is to develop and test a computationally efficient model that can support seasonal planning while preserving key aspects of system operation over hourly and daily timeframes. To do so, an energy system optimization model is repurposed for seasonal planning using features drawn from a unit commitment model. Different scenarios utilizing a well-known test system are used to evaluate the errors associated with both the repurposed energy system model and an imperfect load forecast. The results show that the energy system optimization model using an imperfect load forecast produces differences in monthly cost and generation levels that are less than 2% compared with a unit commitment model using a perfect load forecast. The enhanced energy system optimization model can be solved approximately 100 times faster than the unit commitment model, making it a suitable tool for future work aimed at evaluating seasonal electricity generation and demand under uncertainty.

1. **INTRODUCTION**

Power systems planning has generally focused on two different functions – operations and capacity expansion – which require models that operate at fundamentally different timescales. Unit commitment models (UCMs) are employed for power system operations by considering the hour-by-hour commitment and dispatch of generating units (Wood and Wollenberg, 2012). UCMs assume a time horizon that typically ranges from one day to one week. By contrast, energy system optimization models (ESOMs) are used for capacity expansion planning by considering changes in installed capacity and utilization over future decades. In between these time scales – daily and decadal – an emerging timescale of interest is seasonal (Zhou et al., 2018). Seasonal modeling can potentially lower the cost of electricity supply through improved planning related to seasonal generation and transmission system forced and unforced outages, emissions allowances in coming months, forward purchases of fuel reserves (e.g., coal stock piles), demand response, and hydroelectric releases. Renewable resource availability (e.g., wind, solar insolation, water inflow) is an increasingly important determinant of system dispatch costs (Perez-Arriaga and Battle, 2012), and its temporal variability affects seasonal planning. Electricity demand also exhibits temporal variability at the seasonal scale due to varying temperatures (e.g., Apadula et al., 2012).

At seasonal to interannual time scales, energy demand primarily depends on temperature and is lowest if the mean daily temperature ranges from 60°F-70°F (Changnon et al., 1995). Residential and commercial demands are quite temperature sensitive and significant deviations in mean daily temperatures in a given season can translate into 5-10% fluctuations in total electricity demand, which can severely stress the power grid (Changnon et al., 1995). Thus, both supply and demand of power systems are significantly affected by seasonal variations in climate, which could be predicted in advance based on known climatic and land-surface conditions influencing the region (Oludhe et al., 2013).

Recent advances in monthly-to-seasonal climate predictions show that the skill in predicting both precipitation (Li et al., 2008) and temperature could be utilized to develop hydro inflow and electricity demand forecasts. For example, Changnon and Kunkel (1999) describe the use of climate data and predictions to inform agriculture and water resources management in several applications. This improved monthly-to-seasonal information is also crucial to advise planning and operations in power systems problems such as the hydro-thermal coordination (de Queiroz, 2016). However, there are unavoidable forecast errors in any model, and as a result, real-life power system operations do not precisely match the plan (Jiang et al., 2018). There are different methods for creating electricity demand forecasts based on quantitative techniques (e.g., semi-parametric additive models, autoregressive and moving average models, and exponential smoothing models) and artificial intelligence techniques (e.g., artificial neural networks, fuzzy regression models and support vector machines). While beyond the scope of this paper, Hong and Fan (2016) review the benefits and drawbacks of these methods.

Power system planning at a monthly to seasonal timescale requires electricity demand forecasts, which depend on climate forecasts. But planning at this time scale requires a power system model that incorporates features of both UCMs and ESOMs. For instance, monthly generation levels can be forecast with UCMs by aggregating hourly dispatch decisions. However, the mixed integer linear programing (MILP) formulation associated with UCMs can make them computationally intractable over the monthly to seasonal time scale, especially when embedding uncertainty within the model formulation (Takriti et al., 1996).

By contrast, ESOMs usually have a linear formulation and are computationally tractable but have a coarse grain temporal resolution that limits the accuracy of dispatch decisions, often ignoring the detailed congestion and operational constraints that are explicitly incorporated in UCMs. Detailed consideration of short-term supply and demand is not the core focus of ESOMs (Welsch et al., 2014b). However, in the recent past ESOMs have been adapted and combined with other models to address short-term operational issues with different levels of time discretization and problem features (Koltsaklis and Geordiadis, 2015). For example, Collins et al. (2017) review various methods to capture operational details in ESOMs. Welsch et al. (2014a) demonstrate the need for increased flexibility considerations in long-term ESOMs to more adequately assess future capacity expansion. Welsch et al. (2014b) integrate selected operational constraints (e.g., upward and downward capacity reserve requirements, minimum up and down times, and start-up costs) into OSeMOSYS while maintaining a multidecadal time horizon for capacity expansion in Ireland. Deane et al. (2012) soft-link a UCM (PLEXOS) and an ESOM (TIMES) through combined simulations. By contrast, Kannan and Turton (2013) develop a Swiss TIMES model with 288 time slices but no additional operational constraints, and find the increased temporal resolution more accurately represents the dispatch of variable renewables and flexible gas generation.

Given the gap in timescales covered by UCMs and ESOMs, there is a lack of modeling frameworks that can address power system planning at the seasonal level. This paper fills the gap by repurposing an ESOM to operate at a seasonal timescale. This is the first attempt to reformulate an ESOM in order to create a computationally efficient model that will be able to support seasonal planning while maintaining important finer grain aspects associated with supply and demand over hourly and daily timeframes. Results from the modified ESOM are compared with an existing UCM for validation purposes. In this analysis, Tools for Energy Model Optimization and Analysis (Temoa) (Hunter et al., 2013) is used as the ESOM. A GAMS model (Pandzic et al., 2016b) is used for the UCM. Considering the UCM power dispatch at the monthly time scale as the truth, the errors resulting from both model structure and demand uncertainty are quantified. The proposed approach is tested using the classical IEEE-24 bus system, which has been successfully used in several other analyses, such as hydro-thermal scheduling (Al-Agtash, 2001). In addition, the same test system was used to investigate optimal placement of

  3

energy storage in a power system with a high penetration of wind power (Ghofrani et al., 2013). Kia et al. (2017) use the IEEE-24 bus system to perform day-ahead scheduling of combined heat and power considering thermal storage.

The eventual goal is to use climate forecasts to develop an ensemble of scenarios that include variation in electricity demand and renewable resource supply. These scenarios can be embedded within a single event tree and solved with stochastic optimization to develop a near-term strategy that hedges against different seasonal forecasts. In such a case, the computational performance of the stochastic model becomes a critical issue. Previous work with UCMs includes consideration of day-ahead uncertainty associated with wind power generation using a two-stage stochastic programming approach (Uçkun et al., 2015). Likewise, using an interval optimization approach, Pandzic et al. (2016a) address wind uncertainty in a day-ahead UCM. In addition, stochastic optimization has been applied to uncertainty associated with renewable supply and demand, such as hydro-thermal coordination (Pereira et al., 1991). For example, Silva et al. (2014) perform stochastic optimization to analyze complementarity between wind and hydro power. Jiang et al. (2017) use stochastic optimization for day-ahead dispatch scheduling. Deane et al. (2013) use stochastic optimization to define operational strategies for pumped-hydro storage systems. However, none of these models operate at the seasonal timescale with a precise representation of system operational characteristics and decisions at the hourly level. Given its computational tractability, a modified ESOM that can accurately estimate the hourly dispatch aggregated to the monthly level can be used to perform this stochastic optimization.

This manuscript is organized as follows: Section 2 presents a brief overview of ESOMs and UCMs and points out their key features and differences. Section 3 presents the modeling approach, including enhancements to the ESOM formulation and the demand forecasts used to carry out the comparative analysis. Section 4 presents the simulation results along with a discussion of the case study under different conditions, and Section 5 provides conclusions.

## 2. OVERVIEW OF ENERGY SYSTEM OPTIMIZATION AND UNIT COMMITMENT MODELS

Numerous mathematical programming models focused on the electric sector have been applied to optimize resources and minimize total operational costs over varying time horizons. Hobbs (1995) provides a comprehensive survey of modeling techniques developed for utility resource planning at different time scales and points out the importance of representing uncertainty in such analysis. Oree et al. (2017) review the use of several ESOMs and UCMs in the context of renewable integration challenges and discuss future research directions for modeling power system capacity expansion. ESOMs and UCMs have been developed using various mathematical modeling techniques, including linear programming, MILP, nonlinear programming, and dynamic programming. UCMs generally employ an MILP formulation and consider operational issues with an hourly to weekly timescale, while ESOMs employ a linear programming formulation and consider capacity expansion over multiple decades, with only a coarse-grained representation of supply and demand across different seasons and times-of-day.

*2.1. Unit Commitment Model Characteristics*

UCMs are used to determine the hourly commitment of generators in order to satisfy electricity demand while minimizing startup costs, variable costs associated with fuel usage and operations and maintenance, and fixed costs incurred when the generator is running (Kerr et al., 1966). Baldick (1995) describes classical approaches for solving generalized UCMs. Bard (1988) describes the use of Lagrangian relaxation to solve UCMs. More generally, Sheble and Fahd (1994) provide an overview of the UCM literature. Models designed for this purpose are focused on the electric sector and constructed to investigate problems related to day-ahead market clearing, reliability assessment, intra-day operations, and generation optimal bidding strategies (Zheng et al., 2015). UCMs determine the commitment of each generating unit at each time stage (hours or minutes) in order to minimize the total cost to supply electricity demand while satisfying operational constraints such as load balance, capacity and ramping limits for generators, reliability requirements, and spinning reserves. Several commercially available UCMs can be found, such as GTMax, PLEXOS, and Grid View, as well as academic models, such as Pandzic et al. (2016b), which is employed in this paper. Compared to ESOMs, UCMs represent electricity supply and demand with higher temporal resolution and handle additional operational constraints.

UCMs are typically designed to solve problems for the day- or week-ahead with hourly or intra-hourly discretization of demand (Wang et al., 2013). UCMs are less commonly used to solve problems for horizons spanning one week to one month. Frequently, multi-week problems are modeled as sequential day ahead UCMs, which reduce solution time but do not address operational planning considerations with impacts beyond a few days. Due to the need to model generator commitment, generally represented by binary decision variables (1-on, 0-off), and the structure of the constraint matrix, UCMs designed as MILP programs belong to the class of NP-hard and NP-complete problems (Guan et al., 2003). Because of computational tractability issues, the use of detailed UCMs have mostly been restricted to the development of commitment plans for day- to week-ahead problems.

At a particular time stage $t$, binary decisions have to be made regarding the commitment of the generation units (on/off), startup and shutdown, ramp-up and ramp-down, minimum up and down times as well as continuous operational decisions regarding the physical utilization of generators and transmission lines (Padhy, 2004). The transmission network is represented to consider power flow among the different generation and demand buses. Power balance constraints are represented at each bus of the network, such that the sum of electricity produced by generators connected to a specific bus plus the power flowing into that bus minus the power flowing out from that bus has to be sufficient to meet demand. Ramp up and ramp down limits, startup and shutdown, minimum up and down times for each generation unit are considered structural constraints within UCMs. Operational bounds on decision variables are used to represent system characteristics, such as transmission line capacities, maximum and minimum generation levels, and voltage angles. An outline of a general UCM represented as an MILP is presented in Appendix A.

*2.2. Energy System Optimization Model Characteristics*

ESOMs represent the energy system as a network flow model with multiple technologies linked together by commodity flows. The main goal is to satisfy end-use demands (e.g., vehicle miles traveled, space heating and cooling demand) by making optimal, technology-specific investment and utilization decisions over the model time horizon that minimize the system-wide cost of energy supply. ESOMs address long-term capacity expansion problems over multiple decades, and several different models exist. The MARKAL model (Fishbone and Abilock, 1981) is one of the earliest ESOM representations. The TIMES model is a descendent of MARKAL (Loulou and Labriet, 2007). OSeMOSYS (Howells et al., 2011) is an open source model that has been widely applied in recent years. In these ESOMs, the modeled costs include capital costs for new technologies, fuel costs, as well as fixed and variable maintenance costs. ESOMs include constraints to represent the supply-demand balance, commodity flow through the network, and physical limitations associated with different energy technologies. A general mathematical formulation for an ESOM is provided in Appendix A, and a more detailed model-specific formulation can be found in Hunter et al. (2013).

Generally, these models represent power generation dispatch with a coarse temporal resolution. ESOMs typically group multiple years into a single time period (e.g., 5 years), and the complete set of time periods constitute the model time horizon. The model optimizes a representative year within each time period and assumes that all years within a given time period are identical. To capture diurnal and seasonal variations in energy demand and renewable resource availability, ESOMs split each optimized year into a set of seasons and times-of-day (Welsch et al., 2014a). The flow of energy commodities is balanced for each combination of season and time-of-day, which is referred to as a 'time slice.' It is common practice to represent a limited number of time slices; for example, the default configuration in the MARKAL model generator is two times-of-day (day, night) and three seasons (summer, winter, intermediate) (Loulou et al., 2004). This simplistic representation of supply and demand variation suggests that dispatch results obtained with such models may be sub-optimal.

In this paper, an open source ESOM called 'Temoa' (Hunter et al., 2013) is used, and its formulation is modified to focus on operational decisions rather than capacity expansion. The revised Temoa source code and model data are publicly archived through GitHub (TemoaProject, 2019) and Zenodo (de Queiroz and DeCarolis, 2019). In the tests described below, it is assumed that existing capacity is fixed, and optimal decisions pertain to system operation that satisfies demand at minimum cost.

## 3. METHODS

In this analysis, Temoa is repurposed from a capacity expansion model to an electricity dispatch model, and the results of the dispatch model are compared with the results of a traditional UCM. Figure 1 provides an overview of the modeling process used in this analysis. The left panel represents the model input data pertaining to generators, network topology, costs, electricity demand, and other operational requirements. Information about historical electricity demand is used as input to generate future demand forecasts for each bus in the network. Both the power system characteristics and demand information are used as input to the ESOM and UCM (center panel), which optimize the dispatch of generators over a specific time horizon. As indicated in the right panel, the ESOM and the UCM are both used to quantify differences in monthly electricity generation by plant type and costs across scenarios representing different combinations of demand scenarios and model types.

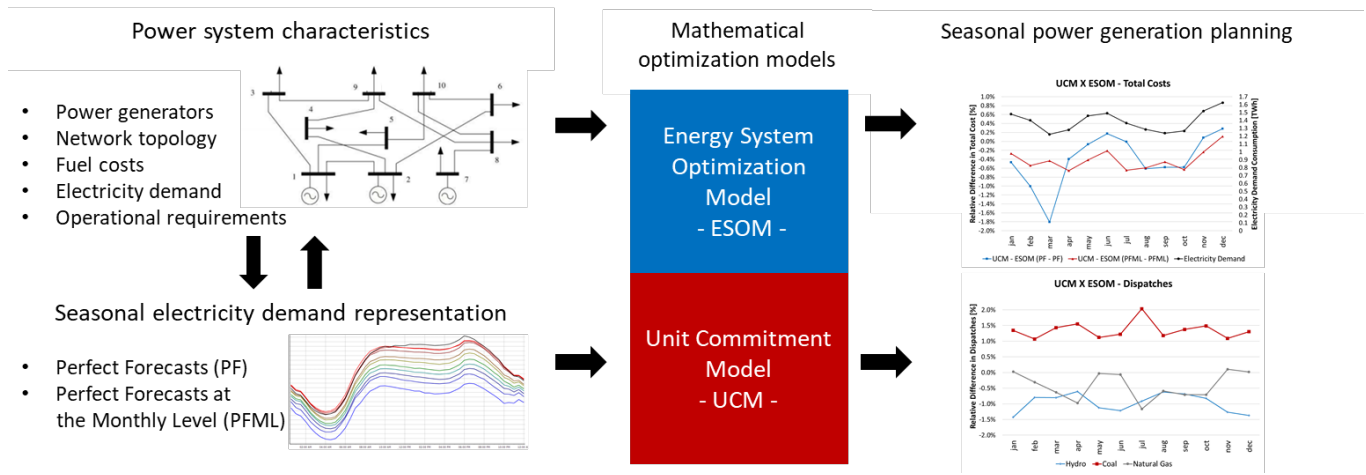

**Figure 1.** Analysis framework for seasonal power generation planning comparison.

### 3.1. *Rolling Horizon Optimization Scheme for Unit Commitment Model*

For comparison with the modified version of Temoa, the UCM model formulation developed by Pandzic et al. (2016b) is implemented with a rolling time horizon. The size of the UCM defined by Equations (A.6)-(A.23) in Appendix A depends on the number of time stages $|T|$, number of generators $|I|$, number of generation block costs $|K|$, number of buses $|S|$ and number of transmission lines $|L|$. However, $|T|$ and $|I|$ define the dimension of the binary decision vectors, and constraints associated with them. For example, in a case with 10 generators and 720 time stages, the model will represent 7200 binary decision variables each for $u_{it}$, $v_{it}$, and $w_{it}$, which represent the generation unit status (on/off), startup status, and shutdown status for each generation unit, respectively. The number of binary decision variables and structural constraints influence the solution time for a large MILP model. Initial attempts to perform UCM runs considering hourly time stages over 1 month were made; however, the majority of the runs were computationally intractable using the CPLEX 12.5 solver and a PC with a 3.4-GHz Intel Core i7 processor and 8-GB 1600-MHz DDR3 memory.

To be able to perform the UCM runs for one month with reasonable computational time (i.e., average solution time of 99.1 minutes), the model is separated into distinct sequential sub-problems (i.e., four weekly sub-problems). Figure 2 shows the flow of initialization data between sub-problems and the subsets of each weekly UCM solution used for the monthly solution. The final monthly solution is based on the combination of the weekly problems depicted by the red vertical lines. The initial conditions for each weekly sub-problem are based on the UCM solution of the previous week. Information passed from one week to the next includes the last dispatch of unit i ($g_{i0}$), the on/off status of each generator i ($u_{i0}$), consecutive hours that operating generators have been online ($UT_{i0}$), and consecutive hours that off-line generators have remained off ($DT_{i0}$). Each week is optimized over nine days (216 hour time periods), and the results of Day 7 are used as the initial conditions for the subsequent week. Simulating Days 8 and 9 ensure a consistent set of decisions in Day 7, but are not used in the final monthly solution.

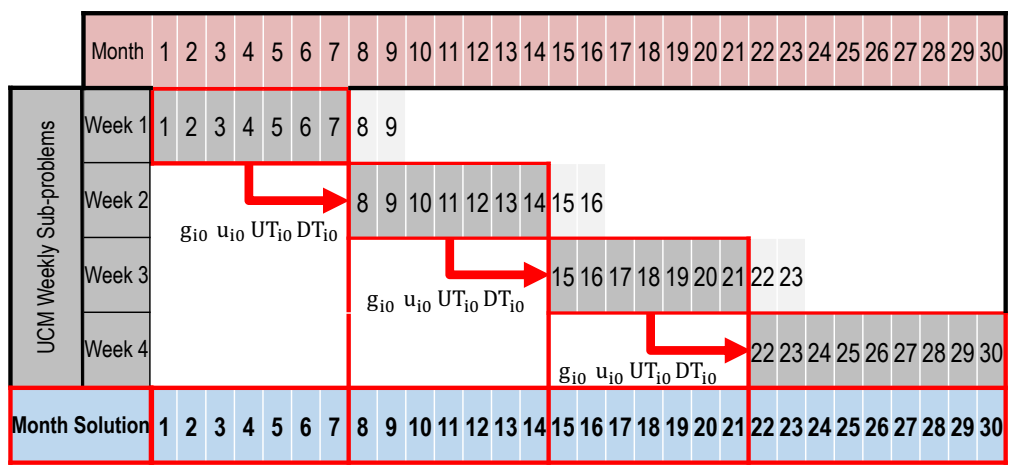

**Figure 2.** The rolling horizon simulation scheme for the unit commitment model.

While this UCM solution is not optimal for the full 30-day horizon, the rolling horizon approach allows for the representation of longer-term problems than the normal day ahead operational planning with a significantly reduced computational time. The approach here is similar to Barrows et al. (2014), where the authors divide simulations into shorter, overlapping periods in order to improve computational tractability without creating large discrepancies in the results. The main advantage of this approach over sequential day ahead solutions is the ability to add constraints that affect intertemporal choices beyond standard day ahead operation planning. Examples of these decisions include when to run hydro generation based on forecasts of reservoir levels or considerations of short-term vs long-term fuel contracts.

While the rolling horizon scheme with the UCM reduces solution time, the computational challenge of using UCMs to solve large horizon problems persist, particularly when considering the potential application of stochastic optimization to account for future uncertainty. For this reason, the goal is to use a modified version of Temoa to capture key operational features of the UCM.

*3.2. Enhancing the Energy System Optimization Model with Unit Commitment Model Features*

Temoa is repurposed by shifting the focus from capacity expansion to power system dispatch. In this case, decision variables related to new capacity installation are disabled, and only operational decisions associated with existing capacity can be optimized in order to minimize cost. In addition, the various time elements in Temoa are remapped such that time periods become months, seasons become days within the month, and times-of-day become individual hours within the day. With this setup, the hourly electricity demand is represented at each bus of the network and has to be met at all times.

In addition, several operational constraints present in the UCM but absent in Temoa are considered:

- Ramp-up and ramp-down constraints (RU/RD), represented by Equations (A.9) and (A.10). These constraints limit the ability of power generation units to increase or decrease their generation output per unit time based on their physical characteristics;
- Minimum up and down times coupled with start-up costs (U/D), represented by Equations (A.17) and (A.18) and the second term in the UCM objective function (A.6). Start-up costs represent the cost to turn on generation units, and the minimum up and down constraints ensure that when a generator changes state (i.e., off to on, or vice versa), it remains in that state for a specified amount of time;
- Fixed costs when the generator is operating (FC), represented by the first term of objective function (A.6). Fixed costs are incurred in each hour that the power generator is operating, regardless of the generation output during that hour.

In modifying Temoa to include these operational constraints, the objective was twofold: avoid the use of integer variables, and keep the model formulation as simple as possible. To help guide the Temoa modifications, several comparative tests are first performed exclusively with the UCM using the IEEE-24 bus test case for unit commitment problems adapted from Diniz (2010). Network topology information for the model is drawn from the IEEE RTS-96 system from Grigg et al. (1996). A description of the IEEE 24-bus case study is presented in Appendix B. Five different UCM instances are created to determine which constraints produce the largest differences in monthly generation with respect to a UCM base case that does not include the RU/RD, U/D and FC constraints. Then, the RU/RD, U/D and FC features are added to the UCM one at a time. The last UCM instance was simulated considering the full model (A.6)-(A.23). The RU/RD and the FC had the most impact on monthly power generation by plant type. The RU/RD constraints have the largest effect on coal generation, which has limited ramping ability. The installed capacity of coal generation corresponds to approximately 35% of the total generation capacity for the system represented in Appendix B. It is important to note that ramping constraints may have less impact in a system with minimal coal generation. To approximate the same results with the ESOM, the revised ESOM formulation with the RU/RD and FC constraints is tested. To incorporate the ramp-up and ramp-down constraints, (A9) and (A10) are added to the set of ESOM constraints. Also, the variable costs ($\rho$) of the power generators are modified in the ESOM to approximate the fixed costs included in the UCM. This variable cost adjustment is an approximation because the true fixed cost is incurred when a generator is up, no matter how much it produces following start up. In the ESOM representation, the fixed costs will be proportional to the electricity produced at a specific hour. After the monthly dispatch was determined, an *ex post* analysis of generators is carried out to compute the true fixed costs incurred by each generator. This approach is chosen to avoid the introduction of binary variables in the ESOM, thereby maintaining computational tractability.

With the enhanced ESOM, results obtained for total cost compared to the UCM were less than 1% over a 1-month horizon; monthly power generation by plant type results are also close to the results obtained by the UCM, as demonstrated in Section 4 where a comprehensive comparison of the enhanced ESOM with the UCM is presented. Other constraints, such as the power flow constraints (A.8) and (A.12), were also tested, but the dispatch results did not vary significantly. However, the observed effects of power flow constraints are related to the current characteristics of the IEEE test system presented in Appendix B. This assumption may need to be revisited in future analyses where transmission bottlenecks are significant. The average computational time for monthly runs of the ESOM is around 60 seconds using similar computer hardware to that used for the UCM runs. In terms of computational software, Python 2.7, Pyomo version 4.3.11388 and CPLEX 12.5 are used to perform the ESOM runs.

*3.3. Demand Modeling: Forecasting Framework and Related Assumptions*

In addition to the challenge of modeling how generators within a given system can meet electricity demand over a given month, future demand itself must be estimated. At the monthly-to-seasonal time scale, forecasting models do not have the ability to precisely forecast hourly or daily electricity demand, hence disaggregation schemes are commonly used to predict hourly and daily demand based on seasonal demand predictions (Prairie et al., 2007). Sinha and Sankarasubramanian (2013) provide additional examples of disaggregation schemes used in streamflow forecasting. Mazrooei et al. (2015) analyze how various disaggregation schemes impact streamflow forecasting over multiple basins across the US Sunbelt. To quantify the utility of the power demand forecast, one could consider actual daily demand, referred to as a perfect forecast (PF), and the daily climatological demand, typically computed over a reference period (e.g., 5-10 years), as two candidate demand forecasts. The skill of any real demand forecast would likely fall between these two extremes. Similar analyses using perfect and climatological forecasts have been used to illustrate how inflow forecasts can improve water supply planning (Sankarasubramanian et al., 2009). Since the purpose is to quantify the effect of demand forecasting errors, two sets of demand scenarios are considered: The first demand representation considers a perfect forecast (PF), with the ESOM and UCM forced to meet observed demand at each hour in each bus of the network for each day of the month during the analysis period. This case is represented as follows:

$$d_{st}^{PF} = \hat{d}_{st}, \qquad \forall s \in S, \forall t \in T \qquad (1)$$

where $s \in S$ is the set of buses of the power system, $t \in T$ is the set of time stages (hours), $d_{st}^{PF}$ is the demand information used in the perfect forecast case; $\hat{d}_{st}$ is the observed demand at bus s and time stage t.

The second demand representation considers a perfect forecast at the monthly level (PFML), i.e., the total electricity demand over each month is considered to be the same as the PF representation, however, the monthly shares allocated to each day and hour are based on the hourly and daily climatological fractions. Relative to using daily climatological demand based purely on historical averages, the selection of this demand forecast method has the purpose to isolate the forecast error associated with sub-monthly demand allocation. For this representation, $m \in M$ is defined as the set of months in the analysis, $n \in N$ as the set of days in a month, and $h \in H$ as the set of hours within a day. Equations (2)-(5) compute the demand in the PFML case at bus s and time stage t ($d_{st}^{PFML}$) for a three-month period (90 days) using one year of historical climate data:

$$d_{st}^{PFML} = I_t \, \bar{f}_h \, \bar{f}_{day} \, \hat{d}_s^m \qquad (2)$$

$$\bar{f}_{day} = \sum_{m=1}^{M} f_{day_m} \Big/ |M|, \quad \forall day = \{1,\dots,30\} \qquad (3)$$

$$f_{day_m} = \frac{\sum_{s \in S} \hat{d}_s^{day_m}}{\sum_{s \in S} \hat{d}_s^m}, \quad \forall m \in M, \forall day = \{1,\dots,30\} \qquad (4)$$

$$\bar{f}_h = \sum_{s \in S} \sum_{n=1}^{|N|} \hat{d}_{shn} \Big/ |N|, \quad \forall h = \{1,\dots 24\} \qquad (5)$$

where, $I_t$ is an indicator function (1 or 0) that maps the information from a specific day during hour h in month m, to a specific time stage t; $\bar{f}_h$ is the climatological average fraction for hour h; $\bar{f}_{day}$ is the climatological average fraction to represent a specific day; $\hat{d}_s^m$ is the total observed demand during month m at bus s; $f_{day_m}$ is the demand daily fraction at a specific day during month m; $\hat{d}_s^{day_m}$ is the observed demand in a specific day in month m at bus s; $\hat{d}_{shn}$ is the observed demand at hour h during day n at bus s. Equation (2) calculates the $d_{st}^{PFML}$ by using the daily climatological average fraction, which is computed using Equations (3) and (4). Equation (5)

   

calculates the climatological average fraction of demand in each hour, which is computed using the observed demand in each hour.

*3.4. Experimental Design for Simulation Comparisons*

A series of comparisons are created to quantify the differences in monthly operational costs and generation by plant type due to (1) model error associated with the revised version of the ESOM compared with the UCM and (2) the demand forecast error. There are a total of four scenarios composed of two models (UCM and ESOM) and two demand scenarios (PF and PFML) considered over 12 months: UCM with Perfect Forecast (UCM-PF), UCM with PF at the monthly level, but imprecise climatological averages at the hourly and daily levels (UCM-PFML), ESOM with Perfect Forecast (ESOM-PF) and ESOM with PF at the monthly level, but imprecise climatological averages at the hourly and daily levels (ESOM-PFML). The UCM-PF is assumed to provide the true power generation levels, since it uses the more detailed UCM with a perfect forecast. Therefore, all the results are shown using the UCM-PF values as the base denominator in the comparison ratios. Three sets of comparisons are carried out (C1, C2 and C3) among the four scenarios:

- Comparison 1: Difference due to model representation under the same demand scenario: Comparison C1a represents UCM-PF × ESOM-PF; and Comparison C1b represents UCM-PFML × ESOM-PFML. The purpose of Comparison 1 is to analyze the difference across models under the same demand representation, where both models use either the PF or PFML demand values. This comparison will inform how much accuracy is lost (in terms of costs and monthly generation levels) when representing the test system using an ESOM instead of the UCM.

- Comparison 2: Difference in model performance due to the different demand scenarios: Comparison C2a represents UCM-PF × UCM-PFML; and Comparison C2b represents ESOM-PF × ESOM-PFML. The purpose of Comparison 2 is to analyze the difference within each model under different demand representations. This comparison quantifies how the monthly generation by plant type and the total costs change in each model (UCM and ESOM) using different daily and hourly demands at each bus of the test system ($d_{st}^{PF}$ and $d_{st}^{PFML}$).

- Comparison 3: Difference due to both model error and imprecise demand: C3 represents UCM-PF × ESOM-PFML. Comparison 3 (C3) quantifies the difference in monthly generation by plant type and the total cost due to differences in both the model (UCM versus ESOM) and the demand forecast (PF versus PFML). In the context of monthly-to-seasonal power generation planning, it is more realistic to assume that imprecise forecasts (instead of perfect hourly values) at a monthly-to-seasonal time scale are available, which could be used as input to the ESOM (instead of the UCM). Therefore, C3 is important in order to understand the compounded effect of using a modified power system model (i.e., ESOM) and imprecise

## 4. RESULTS AND DISCUSSION

The organization of results follows the order of the three comparisons described in Section 3.

*4.1.  Comparison 1: UCM × ESOM using the same electricity demand representations*

Figure 3 presents percentage differences for the UCM × ESOM analysis considering the same demand forecast (PF or PFML) in each comparison. The relative differences are represented in terms of (a) total costs and allocations of (b) hydro, (c) coal, and (d) natural gas plants. Results for nuclear were omitted because they present the exact same values in both models. The total electricity demand (TWh) in each month is also presented in Figure 3a.

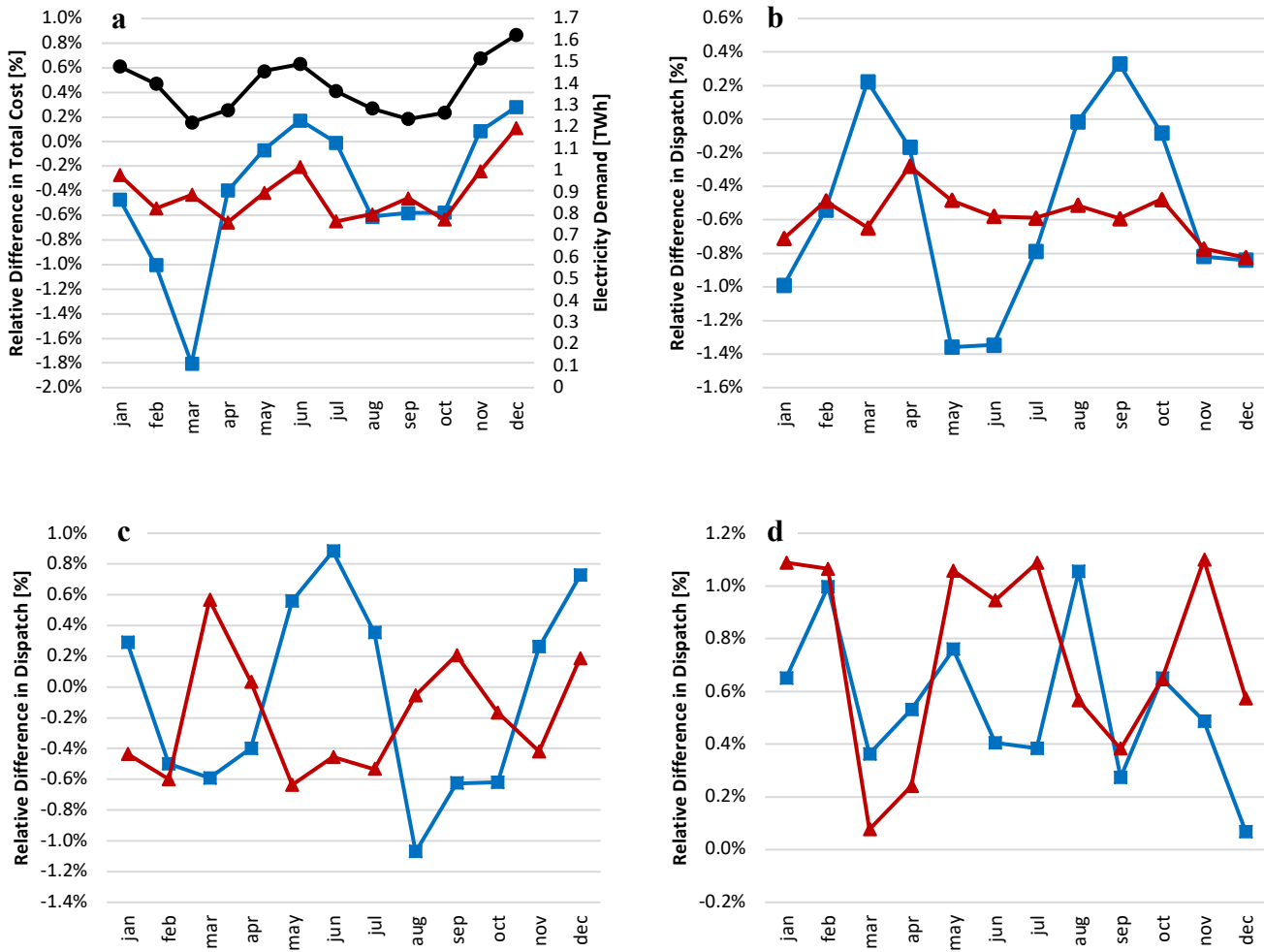

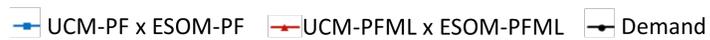

**Figure 3.** Comparison 1 results: UCM and ESOM results with the same electricity demand representations. Panel (a) represents the monthly cost differences and the total system electricity demand consumption in TWh (secondary y-axis); (b), (c), and (d) represent the difference in monthly electricity generation from hydro, coal and natural gas plants, respectively. Note that +% represent larger values in the UCM runs, and -% represent larger values in the ESOM runs.

In Figure 3, positive percentages represent larger values in the UCM runs, and negative percentages represent larger values in the ESOM runs. The total dispatch cost differences between the UCM and the ESOM for the PF (PFML) over the analysis period varies from -1.8% to 0.28% (-0.65% to 0.11%). For hydro generation dispatch, the differences between the UCM and ESOM with PF (PFML) vary from -1.4% to 0.33% (-0.83% to -0.28%). These differences in hydro are compensated by natural gas plants, which vary from 0.07% to 1.05% (0.08% to 1.1%), as shown in Figure 3d, and occasionally by coal power plant dispatch, which varies from -1.07% to +0.88% (-0.64% to 0.57%) over the year, as shown in Figure 3c. A key insight from Figure 3 is that the structural differences between the UCM and ESOM with this generation mix do not contribute to a significant difference in monthly operational cost or generation by plant type. Further, the difference under PF is higher than the difference under PFML, with the PFML case exhibiting less variance over the analysis period, which indicates that the structural difference between the models plays less of a role under imprecise demand. Such information is critical in the context of monthly-to-seasonal planning, since it emphasizes the need to address demand uncertainty.

From this analysis, it is important to note that the enhanced ESOM and the rolling-horizon UCM provide similar estimates of monthly electricity generation by plant type. In months with high electricity demand, the UCM total cost tends to be slightly higher than the ESOM due to a larger use of coal resources to meet the peak demand. The UCM includes a binary representation of startup and shutdown for power plants (not represented in the ESOM), thus, the UCM may choose to keep coal running longer to avoid incurring the fixed startup and shutdown cost. By contrast, the ESOM tends to use more natural gas over the year.

*4.2.    Comparison 2: Standalone analysis of the UCM and ESOM using different demand representations*

Figure 4 presents percentage differences for both models in terms of (a) total costs and the monthly generation from (b) hydro, (c) coal, and (d) natural gas plants. Positive percentages represent a larger value in the PF case, and negative percentages represent larger values in the PFML case. These differences in total monthly generation obtained for the UCM and the ESOM indicate similar generation portfolios under PF and PFML. It is worth noting that the PF run considers the correct representation of peak demand, and that is why the results show a consistently higher cost than the results of the runs using the PFML demand. For the UCM (ESOM), the differences over the year are range from 0.15% to 1.32% (0.31% to 1.52%). The difference in cost is positive and small (less than 2%) between PF and PFML for both models, indicating that the imprecise demand underestimates the true cost, since the coal dispatch is underestimated under PFML (Figure 4c). The difference in cost between the two demand scenarios is also small for both models when the demand is lower than the monthly average.

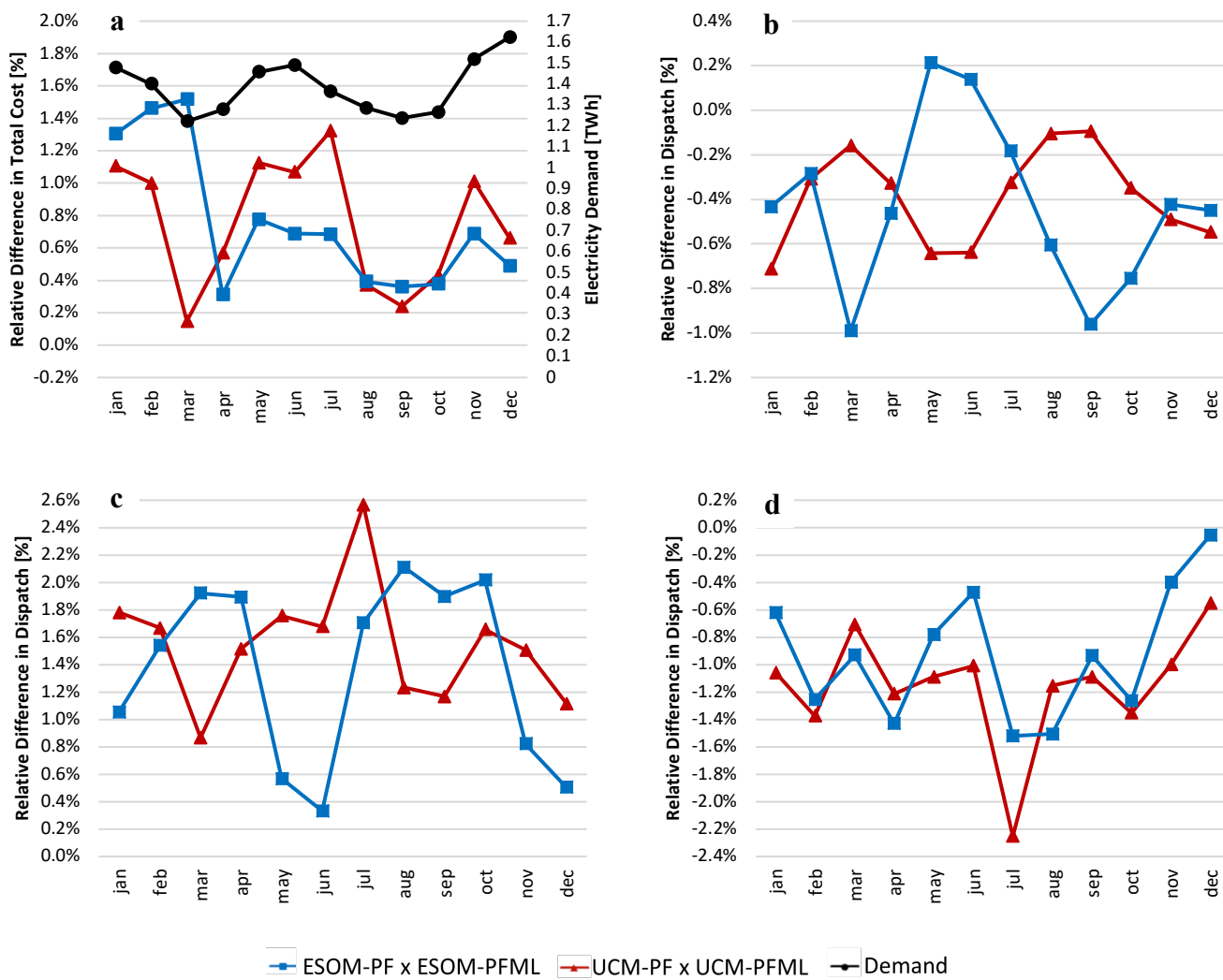

**Figure 4.** Comparison 2 results: stand-alone analysis of the ESOM and UCM for different demand representations over one year. Panel (a) represents the cost differences and the total system electricity demand in TWh (secondary y-axis); (b), (c), and (d) represent monthly electricity generation from hydro, coal and natural gas plants respectively. Note that +% represent larger values in the PF runs, and -% represent larger values in the PFML runs.

In terms of hydro generation, the UCM (ESOM) results range from -0.71% to 0.09% (-0.99% to 0.21%) when using PFML information in comparison with the PF values. Coal and natural gas generation using the UCM (ESOM) range from 0.87% to 2.57% (0.33% to 2.11%) and -2.25% to -0.71 (-1.52% to -0.05%), respectively. In this comparison, both models under the PFML scenario have lower total costs and less electricity generation from natural gas in each month over the twelve-month period compared to the PF runs. These smaller differences in the PFML case can be attributed to less daily variation, since the daily peaks represent the climatological average over the month in the PFML case. However, the percentage differences in costs observed in this comparison, involving different demand representations within the same model, are mostly larger and present a higher variance than the differences observed in Comparison 1. With respect to generation, approximately 2% more coal usage can be observed in the ESOM-PF in Mar-Apr and Aug-Oct when the system demand is at its lowest values, compared to the ESOM-PFML. Coal is the marginal resource in the ESOM-PF scenario, and the lower demand in the PFML scenario therefore reduces coal generation. Smaller differences in coal are observed in the UCM-PFML scenario in comparison with the UCM-PF scenario for the same period. However, the largest differences observed for the UCM are reported in months when the system demand is high. Both models tend to use more natural gas resources in the PFML scenario to account for the difference in coal. Among the resources presented, hydro shows the smallest variation across months.

*4.3.    Comparison 3: Analyzing UCM-PF × ESOM-PFML*

Figure 5 presents percentage differences for the UCM-PF × ESOM-PFML analysis in terms of (a) total costs and (b) monthly generation by plant type. This analysis quantifies the error in applying the ESOM under imprecise demand with the UCM under PF, which represents the perfect scenario. In months with higher demand, the ESOM-PFML costs tend to be slightly lower than the UCM-PF (the largest difference is 0.83%), and in months with lower demand – March, August, September and October – the ESOM-PFML costs tend to be slightly higher than the UCM-PF costs (the largest difference is -0.29%). Overall, the operational cost results in this comparison vary in a small band (from -0.29% to 0.83%) across the 12 months. During the year, the ESOM tends to use more hydropower as well as natural gas, where the UCM tends to use more coal-fired plants. These monthly differences with respect to hydro, natural gas, and coal dispatches range from -1.42% to -0.61%, -1.17 to 0.10%, and 1.0% to 2.0% respectively.

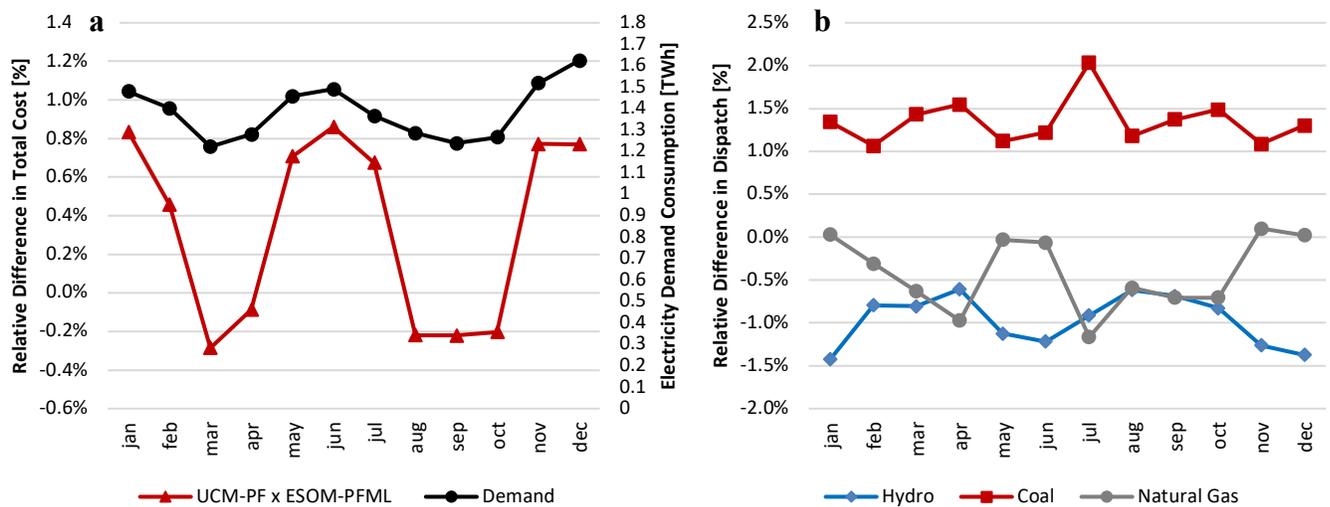

**Figure 5.** Comparison 3 results: monthly comparison of UCM-PF and ESOM-PFML over one year. Panel (a) represents the monthly cost differences (primary y-axis) and the total system electricity demand in TWh (secondary y-axis). Panel (b) represents the monthly generation differences per generation source. Note that +% represent larger values in the UCM-PF runs, and -% represent larger values in the ESOM-PFML runs.

The percentage differences in costs are small throughout the year when the ESOM-PFML is compared with the UCM-PF. The relative differences are larger when the system demand is higher, around 0.8%, due to the representation of demand peaks when the system will need to use more expensive resources to satisfy demand (i.e., peak demand is represented correctly in the UCM-PF, but not in the ESOM-PFML, which uses climatology averages). In terms of electricity generation, the ESOM-PFML uses less coal during the year (around 1.5% less), and this difference is met by a higher usage of natural gas and hydro, which aligns with the lower costs observed with respect to the UCM-PF.

*4.4.    Synthesis – the Energy System Optimization Model versus the Unit Commitment Model*

From the results presented in this section, the error associated with the demand representation plays a larger role in the estimated monthly electricity generation than the differences obtained between the two models under the same forecast. For example, when the models are compared to each other (Comparison 1) annual average cost differences around -0.4% are observed, representing more expensive operation in the ESOM. When the UCM is compared with itself using different demand representations (Comparison 2), higher cost differences are observed, with annual average cost differences around 1.0%. Similar differences are observed for monthly electricity generation. Therefore, in the context of seasonal planning of conventional power systems, the results suggest that the ESOM is a suitable tool to estimate electricity generation months or seasons ahead. Moreover, a more accurate disaggregation procedure of electricity demand at the daily and hourly level will be critical to accurately estimate monthly generation by plant type.

In terms of computational time, the ESOM runs are significantly less expensive than the UCM runs. Each UCM run for one month averaged 99 minutes of computational time, accounting for a total of approximately 40 hours of computational time to perform 24 monthly runs. By contrast, the ESOM runs averaged 1 minute each, accounting for a total of approximately 25 minutes of computational time to perform all 24 monthly runs across both demand scenarios. This performance difference is due to the MILP UCM representation versus the linear ESOM formulation.

Based on the findings of this work, future research should be aimed at expanding the proposed framework to consider seasonal power generation planning in a multi-stage stochastic optimization setting where decisions taken in early time stages may affect future system conditions (e.g., de Queiroz et al., 2019). In such a framework, climate information could be used to inform uncertainty about renewable resource supply. The combination of mathematical optimization models and synthetic demand forecasts can improve operational planning decisions such as forward purchases of fuel, the efficient use water resources, and scheduling plant maintenance. Larger, more realistic power systems with different levels of renewable power penetration and cascading hydropower generation schemes should also be modeled (Jiang et al., 2019). Also, monthly to seasonal energy storage (Hunt et al., 2014) should be analyzed in this context.

## 5. CONCLUSIONS

This paper presents a framework to repurpose an ESOM to perform studies focused on monthly to seasonal power generation planning. An enhanced mathematical formulation of the ESOM is developed to accommodate operational characteristics of the power system, and is applied to an IEEE 24-bus test case. A comparative analysis between the ESOM and a UCM is carried out considering monthly demand disaggregated at the hourly level and represented across a power system network. The study shows that the differences in monthly estimated generation costs and electricity generation by plant type are strongly influenced by the different demand scenarios. Differences in model structure – detailed UCM versus enhanced ESOM – play a minor role in determining monthly electricity generation. Comparative analysis shows that the UCM-PF and ESOM-PFML scenarios produce differences of less than -1.5% to 2% in total monthly cost and generation by plant type (Figure 5). These differences suggest that the ESOM with PFML can provide useful information for monthly-to-seasonal power system planning. In terms of computational efficiency, as noted in Section 4.4, the ESOM is significantly more efficient (approximately 96 times faster), which indicates potential for employing an ESOM for seasonal power generation planning utilizing uncertain demand forecasts. It is important to temper these conclusions by noting that this analysis was carried out using an IEEE test case, and further testing may be warranted. Nonetheless, the results presented here suggest that the modified ESOM can perform large-scale seasonal electricity generation planning analysis with relatively low computational effort and sufficient accuracy.

**Acknowledgements**

 18

This material is based upon work supported by the National Science Foundation under Grant No. CyberSEES-1442909 and the CREDENCE Project (Collaborative Research of Decentralisation, Electrification, Communications and Economics), a US-Ireland Research and Development Partnership Program (centre to centre), funded by The National Science Foundation (0812121), Science Foundation Ireland (16/US-C2C/3290), and the Department for the Economy Northern Ireland (USI 110).

**References**

Al-Agtash, S. "Hydrothermal scheduling by augmented Lagrangian: consideration of transmission constraints and pumped-storage units", IEEE Transactions on power systems, 16(4): 750-756, 2001.

Apadula, F., Bassini, A., Elli, A., and Scapin, S., "Relationships between meteorological variables and monthly electricity demand", Applied Energy, 98, 346-356, 2012.

Baldick, R., "The generalized unit commitment problem." IEEE Transactions on Power Systems 10, 1: 465-475, 1995.

Bard, J. F. "Short-term scheduling of thermal-electric generators using Lagrangian relaxation." Operations Research 36, no. 5: 756-766, 1988.

Barrows, C., Hummon, M., Jones, W., and Hale, E., "Time Domain Partitioning of Electricity Production Cost Simulations," National Renewable Energy Laboratory (NREL), Golden, CO., NREL/TP-6A20-60969, Jan. 2014.

Changnon, S.A., Changnon, J.M., and Changnon, D., "Uses and Applications of Climate Forecasts for Power Utilities", Bulletin of the American Meteorological Society 76(5) 711-720., 1995.

Changnon, S.A. and Kunkel, K.E., "Rapidly Expanding Uses of Climate Data and Information in Agriculture and Water Resources: Causes and Characteristics of New Applications", Bulletin of the American Meteorological Society 80.5 821-830, 1999.

Collins, S., Deane, J.P., Poncelet, K., Panos, E., Pietzcker, R.C., Delaruede, E., Ó Gallachóir, B.P., "Integrating short term variations of the power system into integrated energy system models: A methodological review", Renewable and Sustainable Energy Reviews, 76: 839-856, 2017.

de Queiroz, A.R., "Stochastic hydro-thermal scheduling optimization: An overview", Renewable and Sustainable Energy Reviews, 62: 382-395, 2016.

de Queiroz, A. R., Faria, V. A., Lima, L. M., and Lima, J. W. "Hydropower revenues under the threat of climate change in Brazil", Renewable Energy, 133: 873-882, 2019.

de Queiroz, A.R., DeCarolis, J.F. "TemoaProject: Repurposing an Energy System Optimization Model for Seasonal Power Generation Planning", Zenodo. http://doi.org/10.5281/zenodo.3236502, 2019.

Deane, J.P., Chiodi, A., Gargiulo, M., Ó Gallachóir, B.P., "Soft-linking of a power systems model to an energy systems model", Energy, 42:303–312, 2012.

Deane, J.P., McKeogh, E.J. and O Gallachoir, B.P., "Derivation of intertemporal targets for large pumped hydro energy storage with stochastic optimization", IEEE Transactions on Power Systems, 28(3): 2147-2155, 2013.

Diniz, A.,"Test Cases for Unit Commitment and Hydrothermal Scheduling Problems", IEEE PES General Meeting, 2010.

Fishbone, L.G. and Abilocks, H., "Markal, a linear-programming model for energy systems analysis: technical description of the BNL version", Energy Research, 5:353-375, 1981.

Grigg, C., Wong, P., Albrecht, P., Allan, R., Bhavaraju, M., Billinton, R., et al., "The IEEE Reliability Test System-1996. A report prepared by the Reliability Test System Task Force of the Application of Probability Methods Subcommittee", IEEE Transactions on Power Systems, 14(3): 1010-1020, 1999.

Ghofrani, M., Arabali, A., Etezadi-Amoli, M., and Fadali, M. S., "A framework for optimal placement of energy storage units within a power system with high wind penetration", IEEE Transactions on Sustainable Energy, 4(2), 434-442, 2013.

Grigg, C., Wong, P., Albrecht, P., Allan, R., Bhavaraju, M., Billinton, R., et al., "The IEEE Reliability Test System-1996. A report prepared by the Reliability Test System Task Force of the Application of Probability Methods Subcommittee", IEEE Transactions on Power Systems, 14(3): 1010-1020, 1999.

Guan, X., Zhai, Q., and Papalexopoulos, A., "Optimization based methods for unit commitment: Lagrangian relaxation versus general mixed integer programming", IEEE Power Engineering Society General Meeting, 2003

Hunt, J.D., Freitas, M.A.V., and Junior, A.O.P., "Enhanced-Pumped-Storage: Combining pumped-storage in a yearly storage cycle with dams in cascade in Brazil," Energy, vol. 78, 2014.

Hobbs, B.F., "Optimization methods for electric utility resource planning," European Journal of Operational Research, 83(1):1-20, 1995.

Hong, T., and Fan, S., "Probabilistic electric load forecasting: A tutorial review", International Journal of Forecasting, 32(3), 914-938, 2016.

Howells, M., Rogner, H., Strachan, N., Heaps, C., Huntington, H., Kypreos, S., Hughes, A., Silveira, S., DeCarolis, J., Bazillian, M., and Roehrl, A., "OSeMOSYS: The open source energy modeling system an introduction to its ethos, structure and development", Energy Policy, 39:5850–5870, 2011.

Hunter K., Sreepathi S. and DeCarolis J.F., "Modeling for insight using tools for energy model optimization and analysis (Temoa)" Energy Economics, 40: 339-349, 2013.

Jiang, Z., Li, R., Li, A., and Ji, C., "Runoff forecast uncertainty considered load adjustment model of cascade hydropower stations and its application", Energy, 158: 693-708, 2018.

Jiang, Z., Qin, H., Ji, C., Hu, D., and Zhou, J. "Effect Analysis of Operation Stage Difference on Energy Storage Operation Chart of Cascade Reservoirs", Water Resources Management, 1-17, 2019.

Jiang, Y., Xu, J. Sun, Y., Wei, C., Wang, J., Ke, D., Li, X., Yang, J., Peng, X., and Tang, B., "Day-ahead stochastic economic dispatch of wind integrated power system considering demand response of residential hybrid energy system", Applied Energy, 190: 1126-1137, 2017.

Kannan, R., Turton, H., "A long-term electricity dispatch model with the TIMES framework", Environmental Modeling & Assessment, 18(3): 325-343, 2013.

Kerr, R. H., Scheidt, J. L., Fontanna, A. J., & Wiley, J. K. "Unit commitment", IEEE Transactions on Power Apparatus and Systems, (5), 417-421, 1966.

Kia, M., Nazar, M. S., Sepasian, M. S., Heidari, A., and Siano, P., "Optimal day ahead scheduling of combined heat and power units with electrical and thermal storage considering security constraint of power system", Energy, 120, 241-252, 2017.

Koltsaklis, N. and Georgiadis, M.C., "A multi-period, multi-regional generation expansion planning model incorporating unit commitment constraints", Applied Energy, 158: 310-331, 2015.

Li, S., Goddard, L., and DeWitt, D.G., "Predictive skill of AGCM seasonal climate forecasts subject to different SST prediction methodologies", Journal of Climate, 21(10), 2169–2186, 2008.

Loulou, R., Goldstein, G., Noble, K., "Documentation for the MARKAL family of models", Energy Technology Systems Analysis Programme, OECD/IEA, 2004.

Loulou, R. and Labriet, M., "ETSAP-TIAM: the TIMES integrated assessment model Part I: Model structure", CMS, 5:7–40, 2007.

Ma, J., Silva, V., Belhomme, R., Kirschen, D.S., and Ochoa, L.F., "Evaluating and Planning Flexibility in Sustainable Power Systems," IEEE Transactions on Sustainable Energy, vol. 4, no. 1, pp. 200-209, 2013.

Mazrooei, A., Sinha, T., Sankarasubramanian, A., Kumar, S., Peters-Lidard, C.D., "Decomposition of sources of errors in seasonal streamflow forecasting over the US Sunbelt", Journal of Geophysical Research: Atmospheres, 120(23): 11809-11825, 2015.

Oludhe, C., Sankarasubramanian, A., Sinha, T., Devineni, N. and Lall, U., "The Role of Multimodel Climate Forecasts in Improving Water and Energy Management over the Tana River Basin, Kenya", Journal of Applied Meteorology and Climatology 52(11) 2460-2475, 2013.

Oree, V., Sayed Hassen, S.Z., and Fleming, P.J., "Generation expansion planning optimisation with renewable energy integration: A review," Renewable and Sustainable Energy Reviews, vol. 69, no. Supplement C, 790-803, 2017.

Padhy, N.P., "Unit commitment – a bibliographical survey", IEEE Transactions on Power Systems, 19(2): 1196-1205, 2004.

Pandzic, H., Dvorkin, Y., Qiu, T., Wang, Y., and Kirschen, D, "Toward Cost-Efficient and Reliable Unit Commitment Under Uncertainty", IEEE Transactions on Power Systems, 31(2): 970-982, 2016a.

Pandzic, H., Dvorkin, Y., Qiu, T., Wang, Y., and Kirschen, D, "Unit Commitment under Uncertainty - GAMS Models, Library of the Renewable Energy Analysis Lab (REAL)", University of Washington, Seattle, USA. [Online]. Available at: http://www.ee.washington.edu/research/real/gams_code.html, [Accessed 04.02.16], 2016b.

Pereira, M.V.F. and Pinto, L.M.V.G., "Multi-stage stochastic optimization applied to energy planning", Mathematical Programming, 52:359–375, 1991.

Perez-Arriaga, I. J., and Batlle, C., "Impacts of intermittent renewables on electricity generation system operation", Economics of Energy & Environmental Policy, 1(2), 3-18, 2012.

Prairie, J., Rajagopalan, B., Lall, U. and Fulp, T., "A stochastic nonparametric technique for space-time disaggregation of streamflows" Water Resources Research, 43(3): 1-10, 2007.

Sankarasubramanian, A., Lall, U., Souza Filho, F. A., & Sharma, A., Improved water allocation utilizing probabilistic climate forecasts: Short-term water contracts in a risk management framework. *Water Resources Research*, *45*(11), 2009.

Sheble, G.B., Fahd, G.N., "Unit commitment literature synopysis", IEEE Transactions on Power Systems, 9(1): 128-135, 1994.

Silva, S.R, de Queiroz, A.R., Lima, L.M.M., Lima, J.W.M., "Effects of Wind Penetration in the Scheduling of a Hydro-Dominant Power System", Proceedings of the IEEE Power and Energy Society General Meeting, Washington, 2014.

Sinha, T. and Sankarasubramanian, A., "Role of climate forecasts and initial conditions in developing streamflow and soil moisture forecasts in a rainfall-runoff regime", Hydrology and Earth System Sciences, 17: 721-733, 2013.

Takriti, S., Birge, J.R., and Long, E., "A stochastic model for the unit commitment problem", IEEE Transactions on Power Systems, 11(3): 1497-1508, 1996.

TemoaProject. GitHub: https://github.com/TemoaProject (accessed Sept 13, 2018).

Uçkun, C., Botterud, A. and Birge, J., "An Improved Stochastic Unit Commitment Formulation to Accommodate Wind Uncertainty", IEEE Transactions on Power Systems, vol. 31, 4: 2507-2517, 2015.

Wang, J., Wang, J., Liu, C., and Ruiz, J.P., "Stochastic unit commitment with sub-hourly dispatch constraints", Applied Energy, 105: 418-422, 2013.

Welsch M, Howells M, Hesamzadeh M, Ó Gallachóir B, Deane JP, Strachan N, et al., "Supporting security and adequacy in future energy systems – the need to enhance long-term energy system models to better treat issues related to variability", International Journal of Energy Research, 39: 377-396, 2014a.

Welsch, M., Deane, P., Howells, M., Gallachóir, B.O., Rogan, F., Bazilian, M., and Hans-Holger Rogner, "Incorporating flexibility requirements into long-term energy system models – A case study on high levels of renewable electricity penetration in Ireland", Applied Energy, 135: 600-615, 2014b.

Wood, A. J., & Wollenberg, B. F., "Power generation, operation, and control", John Wiley & Sons, 2012.

Zheng, Q.P., Wang, J. and Liu, A. L., "Stochastic optimization for unit commitment – A review" IEEE Transactions on Power Systems, 30: 1913-1924, 2015.

Zhou, T., Voisin, N., and Fu, T., "Non-stationary hydropower generation projections constrained by environmental and electricity grid operations over the western United States", Environmental Research Letters, 13(7), 074035, 2018.